\documentclass[%
 aps, prd, reprint,
superscriptaddress,
preprintnumbers,
nofootinbib,
 amsmath,amssymb,
 aps,
floatfix,
]{revtex4-2}

\usepackage{graphicx}
\usepackage{dcolumn}
\usepackage{bm}

\usepackage{hyperref}
\usepackage[capitalize]{cleveref}
\usepackage{comment}
\usepackage{subfigure}
\usepackage{xcolor}
\usepackage{soul}
\usepackage{lineno}
\usepackage{upgreek}
\usepackage{endnotes}
\usepackage[normalem]{ulem}

\begin{document}

\title{Total Neutron Cross-section Measurement on CH with a Novel 3D-projection
Scintillator Detector}

\author{A. Agarwal}
\affiliation{University of Pennsylvania, Department of Physics and Astronomy,  Philadelphia, Pennsylvania, U.S.A.}

\author{H. Budd}
\affiliation{University of Rochester, Department of Physics and Astronomy, Rochester, New York, U.S.A.}

\author{J. Cap\'{o}}
\affiliation{Institut de Fisica d'Altes Energies (IFAE), The Barcelona Institute of Science and Technology, Campus UAB, Bellaterra (Barcelona) Spain}

\author{P. Chong}
\affiliation{University of Pennsylvania, Department of Physics and Astronomy,  Philadelphia, Pennsylvania, U.S.A.}

\author{G. Christodoulou}
\affiliation{CERN European Organization for Nuclear Research, CH-1211 Genève 23, Switzerland}

\author{M. Danilov}
\affiliation{Lebedev Physical Institute of the Russian Academy of Sciences, Moscow, Russia}

\author{A. Dergacheva}
\affiliation{Institute for Nuclear Research of the Russian Academy of Sciences, Moscow, Russia}

\author{A. De Roeck}
\affiliation{CERN European Organization for Nuclear Research, CH-1211 Genève 23, Switzerland}

\author{N. Dokania}
\affiliation{Department of Physics and Astronomy, Stony Brook University, Stony Brook, New York, U.S.A.}

\author{D. Douqa}
\affiliation{University of Geneva, Section de Physique, DPNC, Geneva, Switzerland}

\author{K. Dugas}
\affiliation{Louisiana State University, Department of Physics and Astronomy, Baton Rouge, Louisiana, U.S.A.}

\author{S. Fedotov}
\affiliation{Institute for Nuclear Research of the Russian Academy of Sciences, Moscow, Russia}

\author{S. Gwon}
\affiliation{Chung-Ang University, Seoul, South Korea}

\author{R. Howell}
\affiliation{University of Rochester, Department of Physics and Astronomy, Rochester, New York, U.S.A.}

\author{K. Iwamoto}
\affiliation{University of Tokyo, Department of Physics, Tokyo, Japan}

\author{C. Jes\'{u}s-Valls}
\affiliation{Institut de Fisica d'Altes Energies (IFAE), The Barcelona Institute of Science and Technology, Campus UAB, Bellaterra (Barcelona) Spain}

\author{C.K. Jung}
\affiliation{Department of Physics and Astronomy, Stony Brook University, Stony Brook, New York, U.S.A.}

\author{S. P. Kasetti}
\affiliation{Louisiana State University, Department of Physics and Astronomy, Baton Rouge, Louisiana, U.S.A.}

\author{M. Khabibullin}
\affiliation{Institute for Nuclear Research of the Russian Academy of Sciences, Moscow, Russia}

\author{A. Khotjantsev}
\affiliation{Institute for Nuclear Research of the Russian Academy of Sciences, Moscow, Russia}

\author{T. Kikawa}
\affiliation{Kyoto University, Department of Physics, Kyoto, Japan}

\author{U. Kose}
\affiliation{CERN European Organization for Nuclear Research, CH-1211 Genève 23, Switzerland}

\author{Y. Kudenko}
\affiliation{Institute for Nuclear Research of the Russian Academy of Sciences, Moscow, Russia}
\affiliation{Moscow Institute of Engineering and Physics (MEPhl), Moscow, Russia}
\affiliation{Moscow Institute of Physics and Technology (MIPT), Moscow, Russia}

\author{S. Kuribayashi}
\affiliation{Kyoto University, Department of Physics, Kyoto, Japan}

\author{T. Kutter}
\affiliation{Louisiana State University, Department of Physics and Astronomy, Baton Rouge, Louisiana, U.S.A.}

\author{D. Last}
\affiliation{University of Pennsylvania, Department of Physics and Astronomy,  Philadelphia, Pennsylvania, U.S.A.}

\author{L. S. Lin}
\affiliation{University of Pennsylvania, Department of Physics and Astronomy,  Philadelphia, Pennsylvania, U.S.A.}

\author{S. Lin}
\affiliation{Louisiana State University, Department of Physics and Astronomy, Baton Rouge, Louisiana, U.S.A.}

\author{T. Lux}
\affiliation{Institut de Fisica d'Altes Energies (IFAE), The Barcelona Institute of Science and Technology, Campus UAB, Bellaterra (Barcelona) Spain}

\author{S. Manly}
\affiliation{University of Rochester, Department of Physics and Astronomy, Rochester, New York, U.S.A.}

\author{D. A. Martinez Caicedo}
\affiliation{South Dakota School of Mines and Technology, Rapid City, South Dakota, U.S.A.}

\author{S. Martynenko}
\affiliation{Department of Physics and Astronomy, Stony Brook University, Stony Brook, New York, U.S.A.}

\author{T. Matsubara}
\affiliation{High Energy Accelerator Research Organization (KEK), Tsukuba, Ibaraki, Japan}

\author{C. Mauger}
\affiliation{University of Pennsylvania, Department of Physics and Astronomy,  Philadelphia, Pennsylvania, U.S.A.}

\author{K. McFarland}
\affiliation{University of Rochester, Department of Physics and Astronomy, Rochester, New York, U.S.A.}

\author{C. McGrew}
\affiliation{Department of Physics and Astronomy, Stony Brook University, Stony Brook, New York, U.S.A.}

\author{A. Mefodiev}
\affiliation{Institute for Nuclear Research of the Russian Academy of Sciences, Moscow, Russia}

\author{O. Mineev}
\affiliation{Institute for Nuclear Research of the Russian Academy of Sciences, Moscow, Russia}

\author{T. Nakadaira}
\affiliation{High Energy Accelerator Research Organization (KEK), Tsukuba, Ibaraki, Japan}

\author{E. Noah}
\affiliation{University of Geneva, Section de Physique, DPNC, Geneva, Switzerland}

\author{A. Olivier}
\affiliation{University of Rochester, Department of Physics and Astronomy, Rochester, New York, U.S.A.}

\author{V. Paolone}
\affiliation{University of Pittsburgh, Department of Physics and Astronomy, Pittsburgh, Pennsylvania, U.S.A.}

\author{S. Palestini}
\affiliation{CERN European Organization for Nuclear Research, CH-1211 Genève 23, Switzerland}

\author{A. Paul-Torres}
\affiliation{Louisiana State University, Department of Physics and Astronomy, Baton Rouge, Louisiana, U.S.A.}

\author{R. Pellegrino}
\affiliation{University of Pennsylvania, Department of Physics and Astronomy,  Philadelphia, Pennsylvania, U.S.A.}

\author{M.A. Ram\'{i}rez}
\affiliation{University of Pennsylvania, Department of Physics and Astronomy,  Philadelphia, Pennsylvania, U.S.A.}

\author{C. Riccio}
\affiliation{Department of Physics and Astronomy, Stony Brook University, Stony Brook, New York, U.S.A.}

\author{J. Rodriguez Rondon}
\affiliation{South Dakota School of Mines and Technology, Rapid City, South Dakota, U.S.A.}

\author{F. Sanchez}
\affiliation{University of Geneva, Section de Physique, DPNC, Geneva, Switzerland}

\author{D. Sgalaberna}
\affiliation{ETH Zurich, Institute for Particle Physics and Astrophysics, Zurich, Switzerland}

\author{W. Shorrock}\thanks{Now at the University of Sussex, Department of Physics and Astronomy, Brighton}
\affiliation{Imperial College London, Department of Physics, London, United Kingdom}

\author{A. Sitraka}
\affiliation{South Dakota School of Mines and Technology, Rapid City, South Dakota, U.S.A.}

\author{K. Siyeon}
\affiliation{Chung-Ang University, Seoul, South Korea}

\author{N. Skrobova}
\affiliation{Lebedev Physical Institute of the Russian Academy of Sciences, Moscow, Russia}

\author{S. Suvorov}
\affiliation{Institute for Nuclear Research of the Russian Academy of Sciences, Moscow, Russia}

\author{A. Teklu}
\affiliation{Department of Physics and Astronomy, Stony Brook University, Stony Brook, New York, U.S.A.}

\author{M. Tzanov}
\affiliation{Louisiana State University, Department of Physics and Astronomy, Baton Rouge, Louisiana, U.S.A.}

\author{Y. Uchida}
\affiliation{Imperial College London, Department of Physics, London, United Kingdom}

\author{C. Wret}
\affiliation{University of Rochester, Department of Physics and Astronomy, Rochester, New York, U.S.A.}

\author{G. Yang}\thanks{Corresponding author, gyang9@berkeley.edu; Now at the University of California, Berkeley}
\affiliation{Department of Physics and Astronomy, Stony Brook University, Stony Brook, New York, U.S.A.}

\author{N. Yershov}
\affiliation{Institute for Nuclear Research of the Russian Academy of Sciences, Moscow, Russia}

\author{M. Yokoyama}
\affiliation{University of Tokyo, Department of Physics, Tokyo, Japan}

\author{P. Zilberman}
\affiliation{Department of Physics and Astronomy, Stony Brook University, Stony Brook, New York, U.S.A.}

\date{\today}

\begin{abstract}
In order to extract neutrino oscillation parameters, long-baseline neutrino oscillation experiments rely on detailed models of neutrino interactions with nuclei. These models constitute an important source of systematic uncertainty, partially because detectors to date have been blind to final state neutrons. Three-dimensional projection scintillator trackers comprise components of the near detectors of the next generation long-baseline neutrino experiments. Due to the good timing resolution and fine granularity, this technology is capable of measuring neutron kinetic energy in neutrino interactions on an event-by-event basis and will provide valuable data for refining neutrino interaction models and ways to reconstruct neutrino energy. Two prototypes have been exposed to the neutron beamline at Los Alamos National Laboratory (LANL) in both 2019 and 2020, with neutron energies between 0 and 800 MeV. In order to demonstrate the capability of neutron detection, the total neutron-scintillator cross section as a function of neutron energy is measured and compared to external measurements. The measured total neutron cross section in scintillator between 98 and 688 MeV is 0.36 $\pm$ 0.05 barn.

\end{abstract}

\maketitle

\newpage

\section{Introduction}
\label{introduction}

The goal of the current and future long-baseline (LBL) neutrino oscillation experiments is to perform precise measurements of neutrino oscillations, determine the neutrino mass hierarchy and $\theta_{23}$ octant, and test if neutrinos violate CP symmetry. A key element in the measurement sensitivity is the precision with which the neutrino energy of each event can be determined.  

These experiments reconstruct the energy of the neutrino or anti-neutrino based on measurements of the resultant visible particles from the neutrino interaction. 
Neutrons produced in the interactions may carry a significant fraction of the energy and have heretofore been hard to detect and measure. Consequently, neutrons present a significant challenge, but also a major opportunity for improvement in the reliability and precision of the neutrino energy reconstruction. 

\begin{figure}
    \includegraphics[scale=0.1]{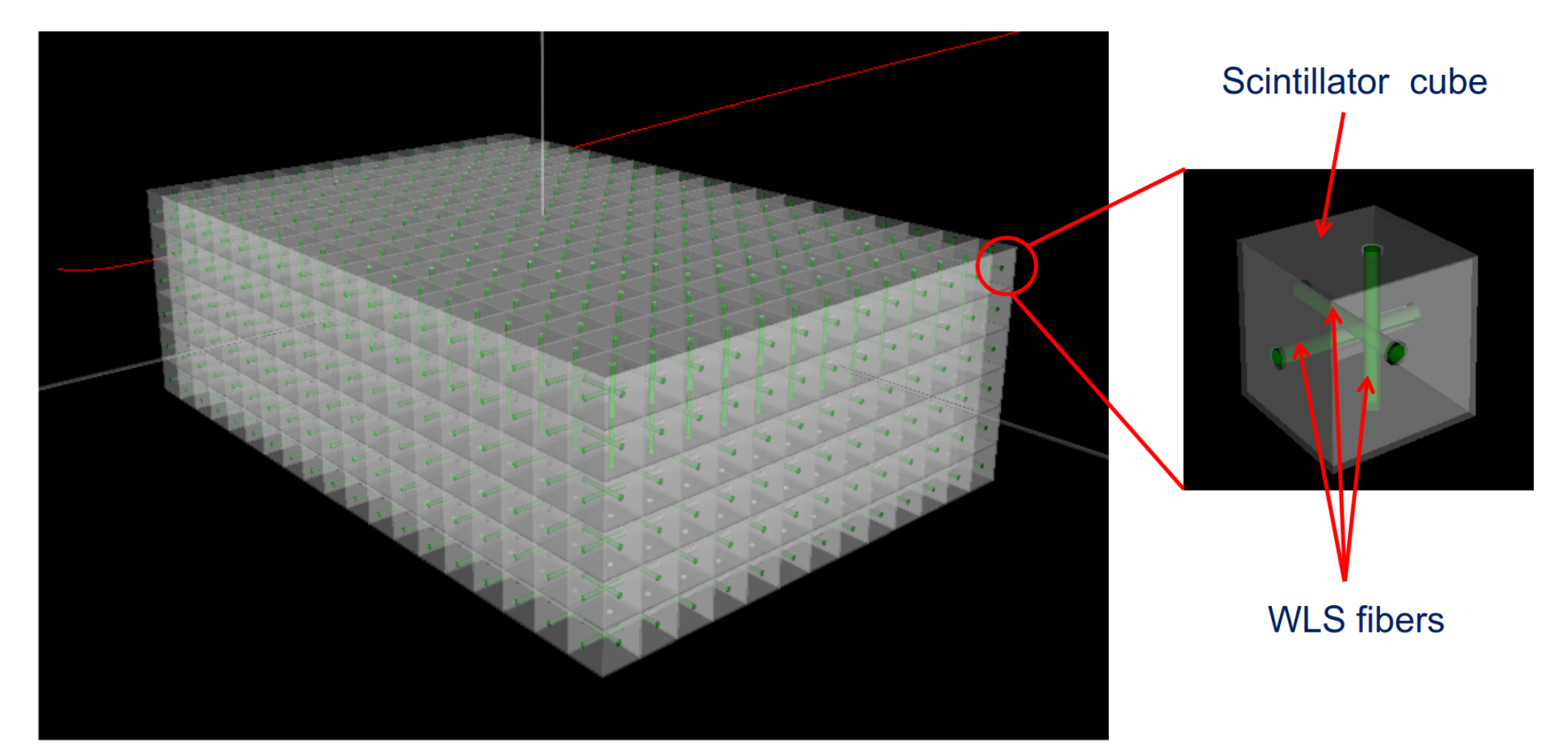}
    \caption{The concept of the 3D-projection scintillator tracker. Figure is taken from~\cite{T2Kupgrade}.}
\label{sfgd_concept}
\end{figure}

The near detectors of LBL experiments must make high-precision measurements of neutrino interactions for a palatable cost. Since weak interaction cross sections are small, the occupancy of neutrino detectors is low compared with detectors measuring charged-particle beams. A novel approach is taken for the upgrade of the T2K near detector~\cite{T2Kupgrade}, employing solid scintillator cubes as the neutrino target. The optically isolated cubes are one centimeter per side and are arranged in a three-dimensional array. The cube surfaces were etched, forming a reflective 50-80 $\mu$m thick polystyrene micropore deposit~\cite{Kudenko:2001qj}. Each cube has three orthogonal holes with a diameter of 1.5 mm. Optical fibers with a diameter of 1 mm pass through the cubes in the $x$, $y$, and $z$ directions (three fibers per cube through these holes). Each fiber passes through a full row or column of cubes. The conceptual design of the detector is shown in \cref{sfgd_concept}.
When a charged particle passes through a cube, light generated by the scintillator is collected by the wavelength-shifting (WLS) optical fibers~\cite{wavelengthcitation}. Photon sensors at the end of each fiber detect this light~\cite{Hamamatsucitation}. Using timing and geometry, the outgoing particles from a neutrino interaction can be reconstructed in three dimensions. This approach yields a significant improvement in reconstruction capability for particle trajectories transverse to the neutrino beam over previous designs which employ planes of scintillator bars~\cite{T2KND280FGD:2012umz}. The approach further enables the measurement of the kinetic energy of outgoing neutrons by time-of-flight (ToF) techniques from energy deposits in the primary medium of the detector. The importance of neutron kinematic detection in the T2K upgrade has been discussed~\cite{T2Kupgrade}~\cite{UpgradeStudy1}~\cite{UpgradeStudy2}.

Our team utilized the Los Alamos National Laboratory (LANL) Weapons Neutron Research (WNR) facility~\cite{Moore1974} to measure the detailed neutron response of two prototypes of the near detector, known as SuperFGD, that will be deployed in the T2K experiment in Japan. We operated two prototype detectors in the beamline for two one-week periods in both December 2019 and December 2020. Here, we describe a measurement of the total neutron cross section on polystyrene using one of the prototype detectors. This measurement represents an improvement in the precision of this cross section for neutron energies between 500 and 688 MeV. 

The paper is arranged as follows. In section II, we discuss the setup of the experiment including the beamline and the detector. In section III, we describe the the detector calibration and in section IV, we introduce the methodology for the total cross-section measurement. In section V, we explain the event reconstruction and selection, followed by the systematic uncertainty consideration in section VI. In the last section, we talk about the measurement result with some discussion.

\section{Experimental setup}
\label{expsetup}

WNR provides a spallation-produced neutron beam with kinetic energies between 0 and 800 MeV. The primary proton beam is composed of sub-nanosecond wide proton bunches separated by 1.8 $\mu$s~\cite{NOWICKI2017374}. Each proton bunch produces photons and neutrons along with other hadrons that are swept away by a magnetic field. The photons and neutrons pass through an aperture in the shielding and traverse the flight path (90 m).  The ToF of neutrons is determined by measuring their arrival times relative to the initial flux of photons (gamma flash). The highest energy neutrons arrive soon after the gamma flash while the lowest energy neutrons come much later, with some arriving after the gamma flash from the subsequent proton bunch (wrap-around neutrons). By positioning our detector at 90 m from the beam target, the farthest location from the tungsten target in the facility, we enhance the energy resolution for the highest energy neutrons while suffering wrap-around for neutrons below 13 MeV. In order to shape the neutron beam profile, a 0.4 cm radius collimator was located 1 m upstream of the detector.

\begin{figure}[h]
\includegraphics[width=1.0\linewidth]{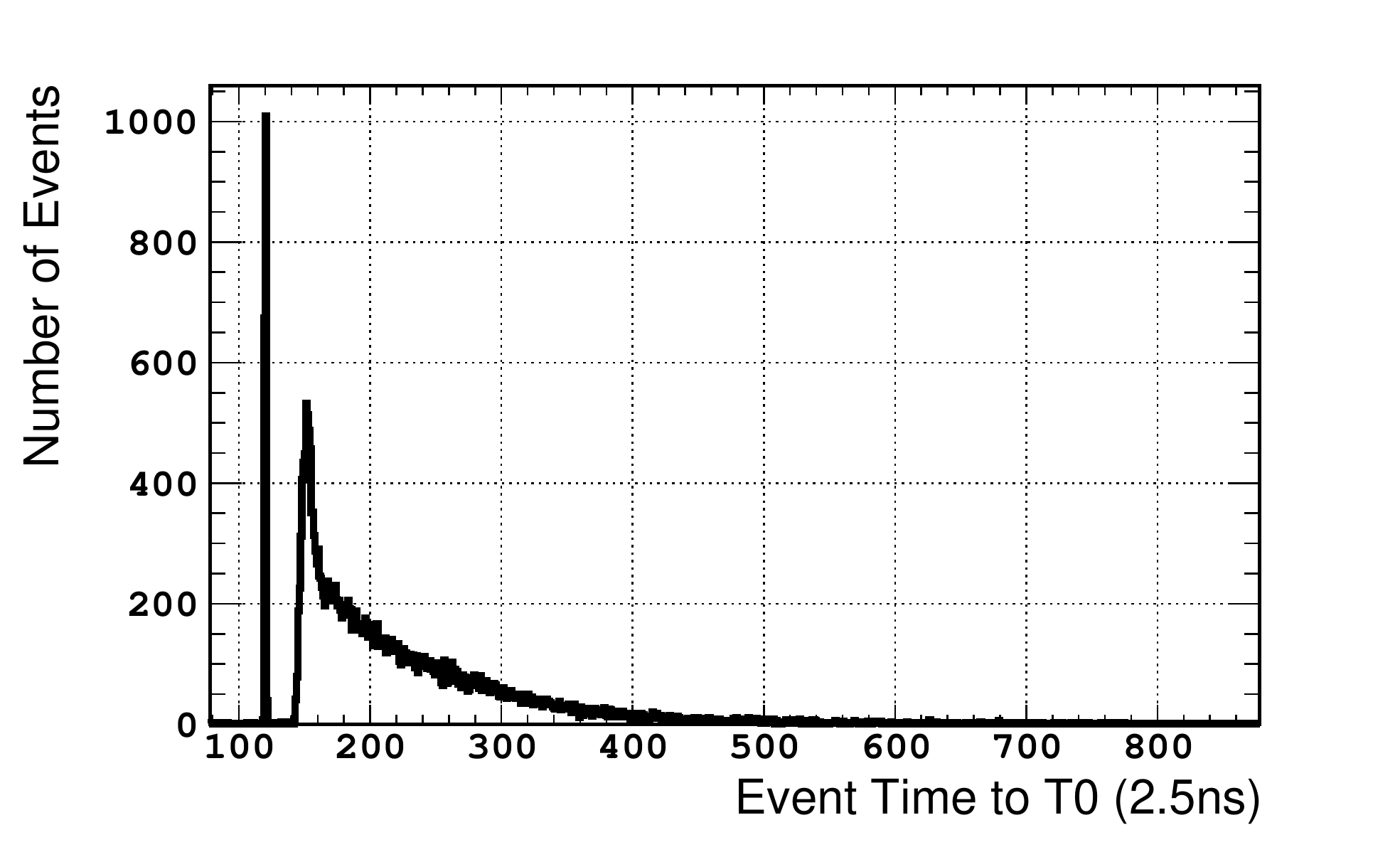}
\caption{Detected event time distribution with respect to $T_0$. The first peak is the gamma flash after the protons hit the tungsten target and the following delayed peak is the neutron. The FEB sampling rate is every 2.5 ns. The binning size is the same as the sampling time size.}
\label{fig:dt}
\end{figure}

The data analyzed here are from our deployment of a 24$\times$8$\times$48 cm$^3$ prototype detector consisting of 9216 1 cm scintillator cubes, exposed to the neutron beam~\cite{Blondel:2017orl}. This is the same detector deployed in a charged particle beam at CERN~\cite{Blondel:2020hml}. The detector was oriented such that the neutron beam ($z$-direction) was parallel to the longest dimension. In the transverse plane, the horizontal ($x$-direction) dimension was 24 cubes wide and the vertical ($y$-direction) dimension was 8 cubes tall. Each energy deposit signal in a cube was collected by three WLS fibers and mapped in three orthogonal views: $XY$ (beam view), $XZ$ (top view) and $YZ$ (side view). The detector used three types of Hamamatsu MPPCs, S13360-1325CS, S13081-050CS and S12571-025C, installed in three different regions of the top view of the detector.  The arrangement can be found in Fig.~5 of Ref.~\cite{Blondel:2020hml}. The beam view and side view are only equipped with Hamamtsu S13360-1325CS MPPCs. The MPPCs signals are read out by customized front-end boards (FEBs)~\cite{BabyMIND}. \cref{fig:dt} shows event time distribution with respect to $T_0$. The $T_0$ is the time proton bunch hits the tungsten target in the beamline. The distribution in \cref{fig:dt} has been obtained using only a subset of the data. The gamma peak width for a single channel, which is 1.4 ns, provides a validation of the timing resolution. A hit is a single fiber channel readout above a threshold. The base threshold for the hit is 40 PE in order to remove the cross-talk. An event includes all hits in a 1.8 $\mu$s time window. The gamma peak appears first in time, followed by the neutron peak. By selecting events in the latter and measuring their time relative to the former, we can determine the energy of the neutron. Reconstructed neutron energy is required to be $> 13$ MeV to get rid of wrap-around neutrons.

\begin{figure}[h]
\centering
\includegraphics[width=1\linewidth]{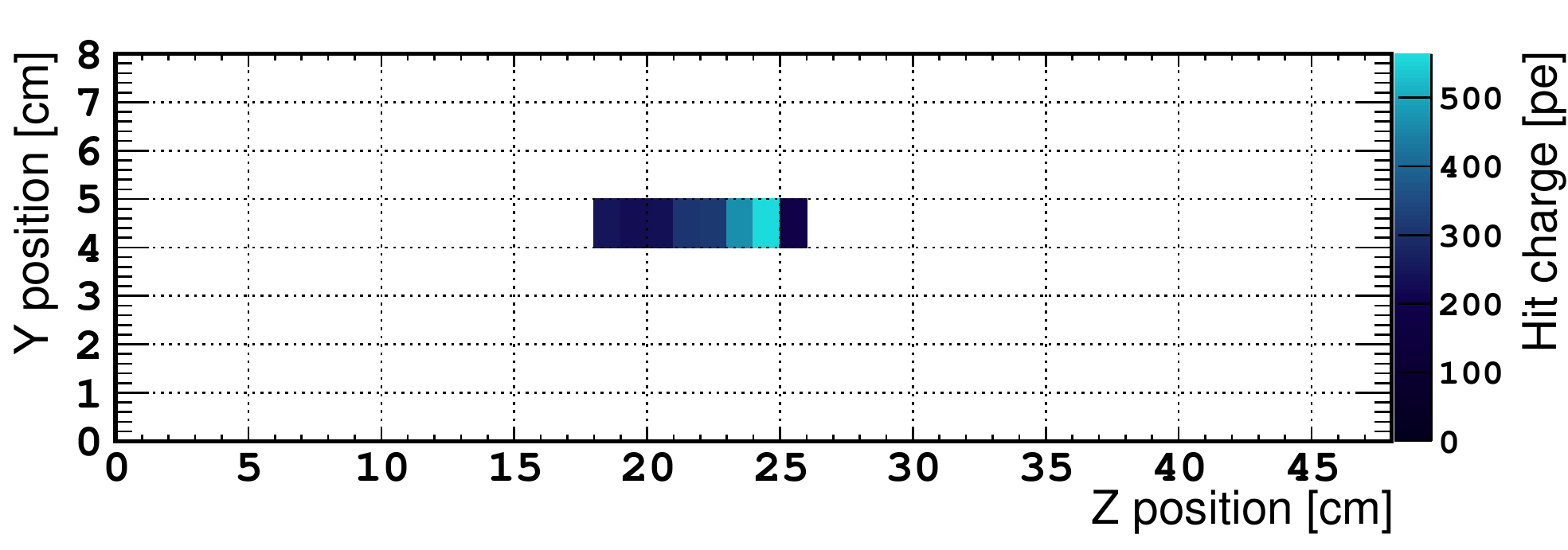}
\caption{Single neutron interaction candidate with 173 MeV TOF-measured kinetic energy on the YZ view. The reconstructed deposited energy is about 112 MeV. Note the Y and Z may have different sizes for one centimeter.}
\label{fig:exampleDisplay}
\end{figure}

\cref{fig:exampleDisplay} shows the hit distributions of a single neutron interaction candidate with a ToF-determined kinetic energy of 173 MeV in the YZ view. The deposited energy is reconstructed to be 112 MeV. For this event the deposited energy is lower than the ToF energy likely because of the binding energy and the invisible particles such as secondary neutrons leaving the detector. 

In 2018, the prototype detector was exposed to a charged particle beam at CERN, where light yield for the minimum ionizing particle (MIP) for one fiber readout was found to be 52.5, 51.6, 42.1 photoelectrons (PE) on average with 8 cm fibers for MPPC type I, II, and III, respectively. The time resolution for a single fiber readout was found to be 1.1 ns~\cite{Blondel:2020hml}.

\section{Calibration}
\label{calibration}

For the gain of a single channel, defined as a fiber readout with a single MPPC and processed through the FEB, the calibration was carried out with a pulsed LED. The gain of each channel was measured in the absence of the beam. The gain is extracted from the PE peaks distribution measured in the presence of LED signals for each MPPC channel using a custom LED system~\cite{customLED,Blondel:2020hml}. 

In addition to the LED calibration, cosmic muons were used to construct a light yield uniformity map for all the detector channels. More than 22,000 through-going muons were selected. The direction of each muon was fitted, thus the light yield per travel length was obtained for the channels that the muon passed through. The average light yield with all muons is used for each channel. Combining the gain and light yield calibrations, a uniform response for all channels was obtained. For the majority of the channels, the light yield measured with cosmic muons is consistent with the expected light yield from the 2018 charged particle beam test. We apply two additional corrections: time walk, the dependence of measured time on the deposited energy, and light attenuation in the fiber (lower than 10$\%$), measured during the CERN beam test~\cite{Blondel:2020hml}.

\section{Methodology for the total cross-section measurement}
\label{crosssection}

The total neutron cross-section measurement was estimated utilizing the so-called extinction technique~\cite{CAPTAIN:2019fxo}. In the presence of neutron-nucleus interactions, the signal event rate decreases exponentially along the $z$-coordinate:

\begin{equation}
	N\left(z\right) = N_{0}e^{-T \sigma_{\mathrm{tot}} z},
	\label{eqxs1}
\end{equation}
where $N_{0}$ and $N\left(z\right)$ are the neutron event rates at the reconstructed $z$ positions in the first layer, and in a more downstream layer in the detector. $T$\footnote{$T=\left(\rho_{\mathrm{CH}}\times N_\mathrm{Avogadro}\right)/m_{\mathrm{CH}}$
$=\,4.623\times10^{22}\,\mathrm{nucleons/cm^{3}}$} and $\sigma_{\mathrm{tot.}}$ are the nuclear density and the neutron total cross section, respectively. Neutron interactions in the detector cause an event rate depletion from which we can extract the neutron total cross section as a function of its kinetic energy. By fitting the event rate distribution along $z$ using an exponential function of the form $N_{0}e^{-\lambda z}$ (in accordance with \cref{eqxs1}), the exponential coefficient $\lambda$, referred to as the extinction coefficient, can be determined. The total cross section is extracted by fitting the event rate distribution along $z$ for each bin of neutron energy. The beam center was measured for each layer to ensure the detector was orthogonal to the beam.

The signal is defined as single-track events because it is easier to identify the vertex and removes the potential issues of pile-up events and light noise. For each layer the cross-section ratio of single-track and multiple-track topologies is constant. Thus, the single-track event rate depletion along $z$ gives the total cross section for each energy range.

In this letter, the measured neutron total cross section on the plastic scintillator (CH), is reported from 98 to 688 MeV. The energy binning was optimized taking into account the energy resolution. The region with low kinetic energy neutron candidates ($<$ 98 MeV) does not result in long enough clusters to form tracks (a few centimeters) and the uncertainty due to 'invisible' scattering\footnote{Invisible scattering includes elastic scattering as well as any interactions that do not produce visible tracks above the threshold.} in this region is large. The region above 688 MeV is statistically limited.

\section{MC simulation}

A realistic geometry has been generated to simulate the detector, experimental hall and the beamline. The finer detector structure such as each cube, cube hole, WLS fiber and the MPPC is implemented in the geometry. Two collimators upstream the detector at 20 m and 89 m have been included in the geometry as well.

The Geant4 simulation is used for neutron interaction and particle propagation in the detector, to provide some of the systematic uncertainty evaluation and model comparison. The \texttt{Bertini} model is chosen as our default model~\cite{WRIGHT2015175}. As an alternative, the \texttt{INCLXX} model is used throughout the analysis to cross-check against any model dependence introduced by the choice of the default model~\cite{PhysRevC.66.044615}. For both the systematic uncertainty evaluation and the total cross section comparison, these two models show very consistent results, thus for the remaining of the paper, we present the results with \texttt{Bertini} model. Our cross-section measurement has no simulation or model dependency.

\begin{figure}[h]
\centering
\includegraphics[width=1\linewidth]{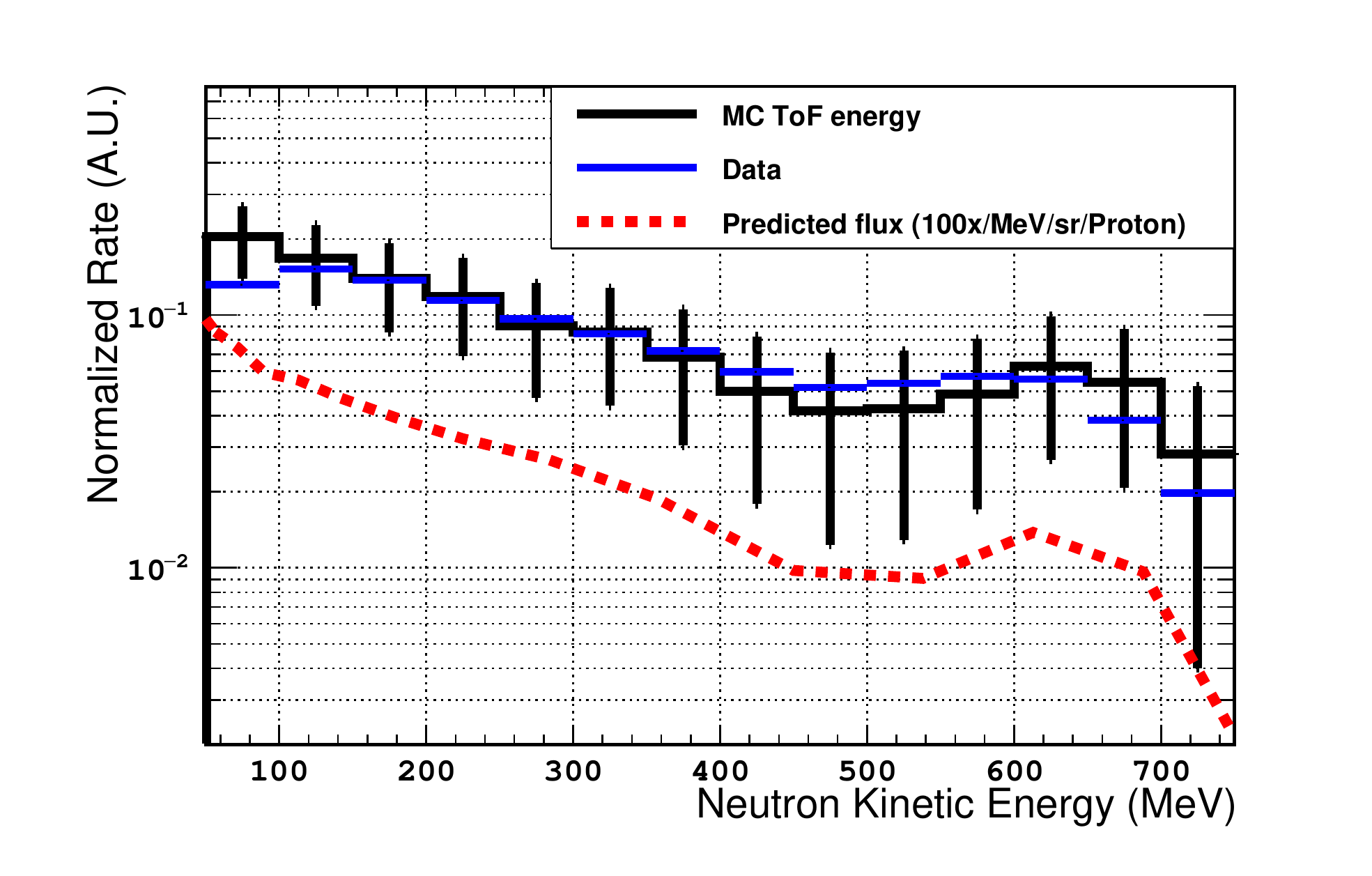}
\caption{Neutron energy spectra obtained with data (blue line) and MC (black line). The error bars represent the MC statistical uncertainty, while the statistical uncertainty associated with data is included but too small to be visible. The measured flux (red dashed line) is included as reference.}
\label{fig:energy}
\end{figure}

A comparison between the reconstructed neutron energy distribution in MC and data is shown in \cref{fig:energy}. The simulated events were generated following the measured neutron flux\cite{flux-private-comm}. The measured neutron energy in data and simulation are consistent within the MC statistical error. It is worth to note that the scintillation light yield non-linearity is not included in the current simulation. Constant light yields extracted from the data are used for the MC simulation. In future studies, this effect will be taken into account.

\section{Event reconstruction and selection}\label{reconstruction}

The goal of the event reconstruction for this analysis is to aggregate information on hits recorded by the prototype and convert them into information from which we select neutron interactions producing one reconstructed track in the event. Only events occuring after the gamma flash are selected, for this reason we select hits collected in a time window between $-815$ and 850 ns with respect to T$_0$. To reject hits produced by MPPC noise and optical cross talks in adjacent cubes, we set a base threshold of 20 PE per hit and we require a minimum number of three hits in an event. After having sorted the hits in time, a time clustering algorithm is used: if consecutive hits are greater than 17.5 ns apart they belong to different clusters. This time clustering value has been optimized such that the probability of overlapping neutrons in one single event is below 0.1$\%$.

After that, only events with one time cluster are selected to avoid pile-up with other neutron interactions. At this point, voxels, defined as reconstructed cubes, are built combining the three views of the detector and a density-based spatial clustering algorithm (DBScan) is employed to select voxels close in space~\cite{Ester96adensity-based}. A cluster is defined if there is at least one voxel and if the distance between voxels does not exceed 1.8 cm, which ensures there are not holes between voxels. Once the spatial clusters are built, we select events with only one spatial cluster. The particle traveling in the beginning and end voxels may not go through the full cube. In order to remove the z-dependency on the selection, layer-dependent PE cuts on the first and last voxels are applied for each track. The PE cut values are determined by minimizing the track length distribution difference among all layers.

For each cluster a matrix is defined that encodes the distance between the center of the cluster and each voxel:  
\begin{equation}
    M_{ij} = \sum^N_{ij} \frac{(\vec{v} - \vec{c})_i (\vec{v} - \vec{c})_j}{N},
\end{equation}
where $N$ is the total number of voxels in the cluster, $\vec{v}$ is the displacement vector associated with a voxel in the cluster and $\vec{c}$ is the center of the cluster. A Principal Component Analysis (PCA) of this matrix is performed to find the three principal eigenvectors and eigenvalues. A variable defined as $L = (\lambda_1 - \lambda_2)/\lambda_1$, where $\lambda_1$ and $\lambda_2$ are the first two eigenvalues of the PCA components, quantifies the development of the cluster along the principal vector. Requiring $L>$ 0.7 rejects clusters inconsistent with a single final state particle origin. 
Two other variables are employed to select straight tracks. The first is the maximum distance between a voxel and the line defined by the largest principal vector. The second is the largest projected distance between two voxels on the second principle eigenvector. They are required to be below 1.2 cm and 1.4 cm, respectively.

Only events having between three and eight voxels are accepted to reduce the dependency on the detector acceptance. For such events, the first voxel along the $z$-coordinate is taken as the vertex. In addition, only events with a vertex within a rectangular parallelepiped built around the beam center of 1.5$\times$1.5$\times$40 cm$^3$ are selected.

The first $z$ layer of the prototype is rejected since it is contaminated by the hits produced by particles from the interactions in the upstream material and collimator. The last nine layers are rejected as a result of the cut on the number of voxels mentioned above. The timing associated with the selected vertex is used to compute the energy of the neutron using ToF.

There is almost no background in the final sample. The overlapping events are below 0.1$\%$ due to the low event rate and they are rejected by the time and space clustering. Multiple interactions are rejected by the single cluster selection. The remaining multiple interactions with the first interaction invisible are included in the invisible scattering uncertainty described in the next section. The background from the neutron interaction in the collimator upstream of the detector and cosmic muons have been rejected by cuts on the fiducial volume. The final sample purity is above 99$\%$.

\section{Systematic uncertainties}

The detector has a geometrical and electronics non-uniformity which generates a detection uncertainty. The fiber, fiber hole, MPPC, insulating Tyvek layer and cubes have some variation in their alignment. For our neutron experiment, the beam profile is rather narrow and our measurements may be sensitive to such variations in the detector. Additionally, we employ three different types of MPPCs for the prototype and they are not uniformly deployed on the detector. 

\begin{figure}[h]
\includegraphics[width=1.0\linewidth]{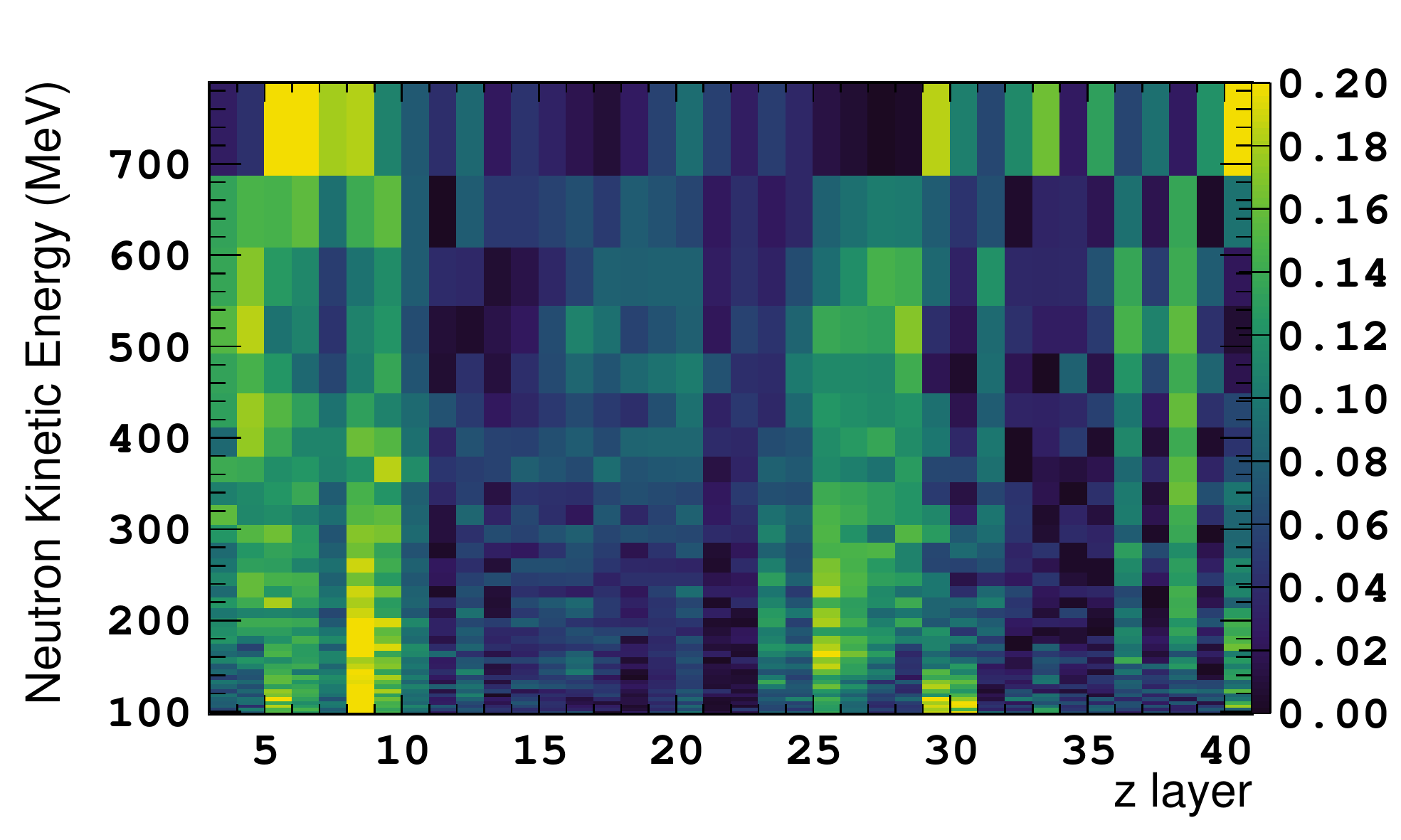}
\caption{The fractional systematic uncertainty due to the detector non-uniformity and reconstruction inefficiency. See text for detail.}
\label{fig:detection2}
\end{figure}

To evaluate the uncertainty associated with these variations, we compare our results to a ``no-cut'' sample. The no-cut sample only requires a hit to have more than 20 PE and any number of reconstructed voxels. All other topological cuts existing in the single-track selection are removed. For each energy range, the event rate along $z$ in the ``no-cut'' sample was normalized to that in the single-track sample. Then the event rate fractional difference at each layer between the two samples is taken as the detector systematic uncertainty. \cref{fig:detection2} shows the resulting systematic uncertainty by calculating the residual of the two samples as a function of the neutron kinetic energy and $z$. The ``no-cut'' sample is normalized to the single-track signal sample. To understand the cause of the detector systematic uncertainty, three dedicated studies were completed. First, we compared the event rate along Z with the same rate in the data taken with the detector rotated by 180 degrees around Y (vertical). The ratio between the two is consistent with the fractional uncertainty shown in~\cref{fig:detection2}. This ruled out that the reconstruction is the major reason for such uncertainty. Second, we used only two views, the beam and side view, which are equipped with the same MPPC type. Also in this case we compared the event rate we measured with the non- and rotated detector finding a good agreement with the estimated fractional uncertainty. This test indicates that the difference between MPPC-type is not the major reason as well. Then, we performed a MC simulation introducing a misalignment of $\pm$0.5 millimeter for groups of 10 cubes. Such misalignment can introduce variations similar to the fractional systematic uncertainty shown in~\cref{fig:detection2}. Therefore, we concluded that the dominant reason for the detector systematic uncertainty is likely to be the cube misalignment due to the fact that the cubes were assembled without a solid support on the bottom, but they were supported only by the fibers.

The extinction method requires the reconstruction of the first interaction vertex. Intrinsic contamination arises when neutrons scatter, leaving a deposited energy below the detector threshold. In this case, the primary vertex is missed and a subsequent scattering may be mis-reconstructed as the vertex. Generally, the secondary vertex is downstream of the true primary vertex. 

This invisible scattering constitutes a source of systematic uncertainty. In order to estimate the uncertainty, we studied both the \texttt{Bertini} and \texttt{INCLXX} models in Geant4 version \texttt{10.3}, and the conclusions are consistent with each other. One of the consequences of invisible scattering is that it generates transverse spread of the neutron beam. We conservatively assumed all of the transverse spread comes from invisible scattering, tuned the simulations to match the spread seen in data, and calculated the change in the reconstructed neutron cross section. 

\begin{figure}[h]
\includegraphics[width=1.0\linewidth]{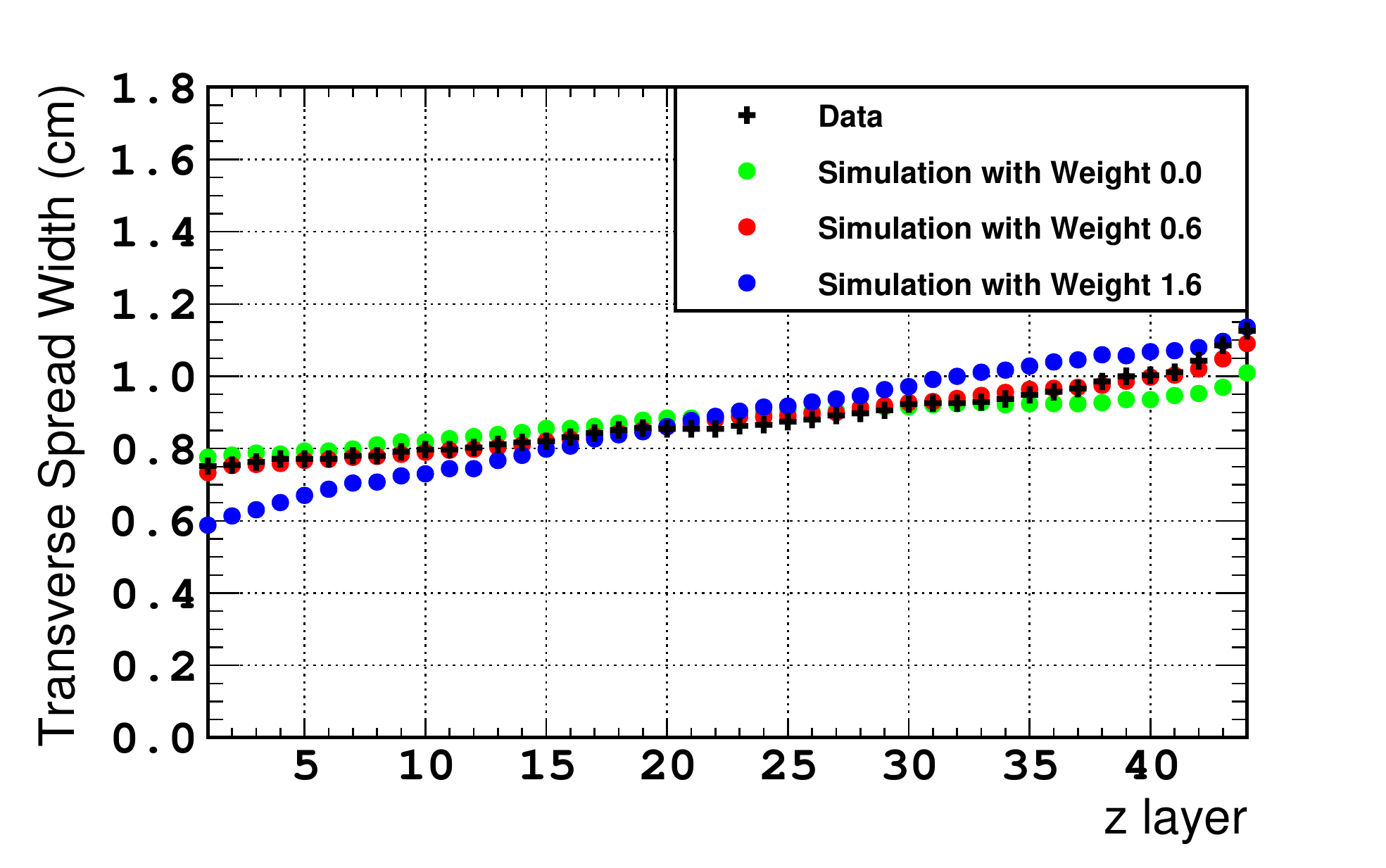}
\caption{Width of the vertex transverse spread as a function of Z with different invisible scattering strength. The weight reported in the legend is the factor applied to the original invisible scattering strength in the model. For example, simulation with weight 0 corresponds to no invisible scattering including elastic scattering. MC simulation has been normalized to data.}
\label{fig:invisible1}
\end{figure}

\cref{fig:invisible1} shows the width of the vertex transverse spread as a function of $z$ with different fractions of invisible scattering in the MC simulation (indicated as "weight" in the legend). Varying such fractions change the strength of invisible scattering. For example, simulation with weight 0 correspond to absence of invisible scattering including elastic scattering. If an event had one invisible scattering before leaving visible signal, the event was assigned a weight of 0.6. If twice, it was assigned a weight of 0.36. The weight of 0.6 shows the best match with data.

The impact of the invisible scattering is as large as 10\% below 100 MeV but limited to a few percent above 100 MeV. The difference between the cross section results with and without the invisible scattering is taken as a systematic uncertainty.

\cref{fig:invisible1} shows the transverse spread as a function of $z$ integrated over the entire neutron energy range. In general, the spread increases as function of the depth in the detector and is more pronounced for low energy than high energy neutrons.

The limited dimensions of the prototype have an impact on the single-track event selection efficiency. For the single-track selection, the downstream part of the detector has higher efficiency than the upstream part. The downstream part reduces the room for multiple-track events to develop. A data-driven correction is considered in order to avoid any possible bias. We compare the number of events selected at a given $z$-layer with different total lengths of fiducial volume (e.g. $z$-layer is 2 and use the detector through $z$-layer 48, then $z$-layer is 2 and use the detector through $z$-layer 47, and so on). The uncertainty on the acceptance correction was computed by setting the starting layer to 2 through 8 and taking the largest acceptance correction difference between any two fixed layers. The resulting uncertainty is under 2\% for the whole energy range.

The finite timing resolution results in an uncertainty on the neutron ToF and consequently its reconstructed kinetic energy. The overall uncertainty was computed combining the resolution for a single fiber mentioned above and the uncertainty on T$_0$, and it was found to be 1.37 ns. 
 
Contribution to the uncertainty on the cross section by the timing uncertainty is estimated varying the ToF values thousands of times according to a Gaussian distribution centered at the actual ToF and a width corresponding to the timing resolution. The cross section is extracted for every variation and the spread of the resulting distribution is taken as the uncertainty. In addition, in order to evaluate the light yield variation, for each channel, the light yield fluctuation estimated by the cosmic muon track fitting is propagated through the reconstruction with a simulated neutron interaction sample, providing the uncertainty on the vertex location induced by light yield fluctuations. 

Finally, we considered uncertainty due to neutrons interacting in the collimators located upstream of the detector. If neutrons lose energy in such interactions, they can arrive in the detector with an energy lower than that reconstructed by ToF. We studied this effect with simulated data and found it to be negligible. Generally, neutrons which lose a large amount of energy in the collimators are scattered transversely and are not included in our data sample. Neutrons that scatter in the collimator and interact in the detector lose less than 1 MeV. This uncertainty is propagated to the cross-section uncertainty by varying the energy distribution according with the spread induced by the neutron interactions inside the collimators.

Contribution by the uncertainties that do not change the neutron energy are computed as a function of $z$-layer and neutron energy (e.g. \cref{fig:detection2}). They are propagated in the following way: the number of events in each $z$-layer and energy bin is varied according to a Gaussian with mean the number of events in that particular bin and width the estimated uncertainty. The neutron cross section is extracted from every variation and the width of the distribution is taken as uncertainty. As stated before, the major cause of the systematic uncertainty is the cube misalignment. The correlation of the cube misalignment at each layer is not assessable. This measurement relies on this assumption and the future experiments should consider dedicated ways to measure the uncertainty induced by the cube misalignment. For example, systematic cube alignment variations can be conducted to understand the impact of the misalignment. 

Additionally, we performed the following checks to confirm the robustness of the analysis:

\begin{itemize}
    \item The cross-section was extracted from a MC sample obtained employing the same reconstruction and selection as for data. The results were consistent with the input total cross section within the uncertainties we have evaluated.
    \item Data were divided into individual calendar days and comparison of vertex distributions between each other showed consistency.
    \item Different fitting ranges were employed to understand the local structure of the $z$ distribution 
    \item A constant term was added into the exponential function to to include any potential constant background such as noise or external background across $z$. It results in a consistent cross-section measurement as the one obtained using the nominal exponential function. 
    \item The total cross section was re-evaluated by refitting the exponential function after subtracting the invisible scattering predicted by the tuned simulation for each $z$ layer from data. The resulting total cross section is consistent with the original result. 
\end{itemize}

\section{Result and discussion} 
\label{results}

\begin{figure}[h!]
\includegraphics[width=1.0\linewidth]{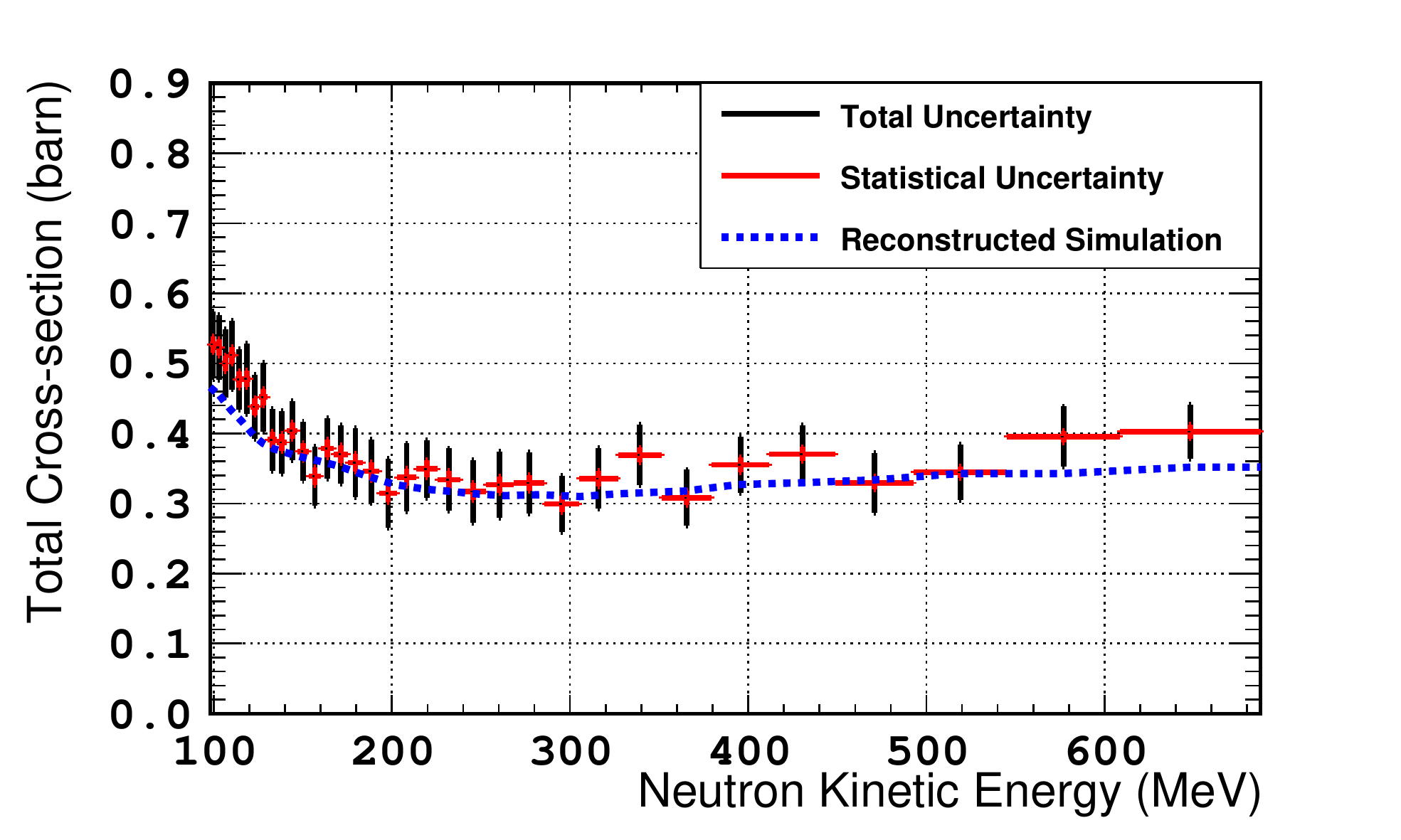}
\caption{The total neutron-CH cross section as a function of neutron kinetic energy. The black vertical bars represent the total uncertainty and the red the statistical uncertainty. The Geant4 \texttt{Bertini} model is shown in blue.}
\label{fig:result1}
\end{figure}

With the exponential fit to the vertex distribution along z for each energy range, the total neutron-CH cross section is obtained as a function of the neutron kinetic energy. All the systematic uncertainties discussed above are included varied separately and then summed in quadrature. The total neutron cross section on hydrocarbon as a function of neutron kinetic energy is shown in~\cref{fig:result1}. For neutron energies below 200 MeV, we obtain a slightly higher measured value than the \texttt{Bertini} model in Geant4.

\begin{figure}[h!]
\includegraphics[width=1.0\linewidth]{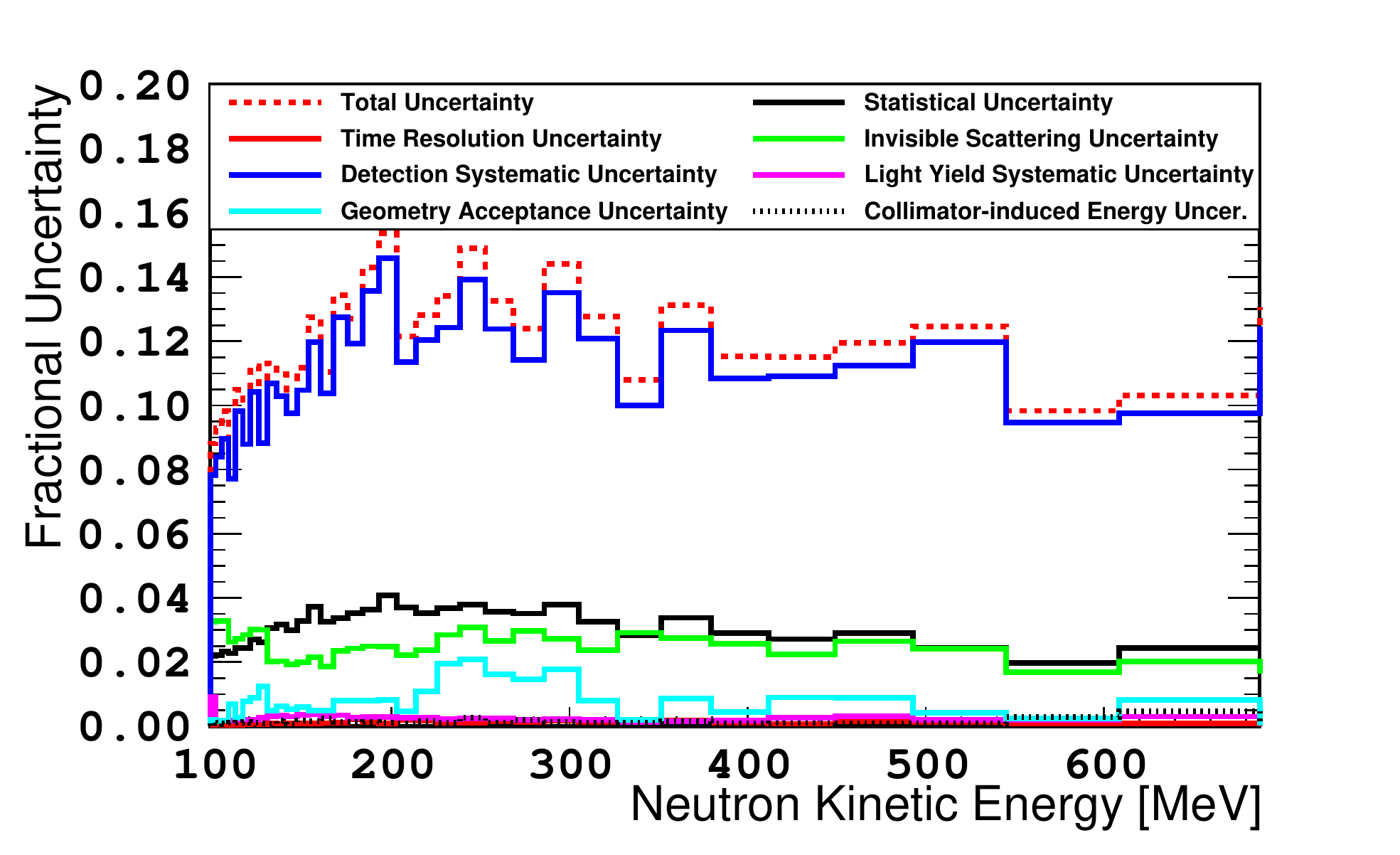}
\caption{Breakdown of the total uncertainty. The systematic uncertainty includes contributions from invisible scattering, light yield, detection and reconstruction, time resolution, collimator interaction and geometric acceptance correction.}
\label{fig:result2}
\end{figure}

In the energy region between 200 to 688 MeV, our measurement shows good agreement with the model. The overall data and the model agree within the uncertainties. It should be noted that for each energy bin, the measurement is effectively independent and systematic uncertainties are considered uncorrelated across energy bins. For the region below 150 MeV, the $\chi^{2}$/d.o.f. is 15.2/11 and for that below 200 MeV, the $\chi^{2}$/d.o.f. is 16.1/18. The energy-integrated (98-688 MeV) cross section is 0.36 $\pm$ 0.05 barn with a $\chi^{2}/d.o.f.$ of 22.03/38, corresponding to 38 layers in $z$. The comparison between the true and reconstructed cross sections in the MC has been done in order to understand the bias introduced by the reconstruction. The difference among them is well within the error bars. The total uncertainty is broken down into each contribution by various sources in \cref{fig:result2}. The total uncertainty is dominated by the contribution from the detection systematic uncertainty. 

Our result on CH from 98 to about 600 MeV is consistent within the error bar with existing neutron-Carbon total cross-section measurements\cite{Schimmerling:1973bb, Dimbylow_1980, Finlay:1993hk, Abfalterer:2001gw}, indicating a minor contribution from the neutron-Hydrogen interaction. Our measurement provides a new result on the total neutron-CH cross section across a broad energy range that is important for the LBL neutrino oscillation experiments.

\section{Data availability} 

Data will be made available on request.

\begin{acknowledgments}

This work was performed, in part, at the Los Alamos Neutron Science Center (LANSCE), a NNSA User Facility operated for the U.S. Department of Energy (DOE) by Los Alamos National Laboratory (Contract 89233218CNA000001). We thank Keegan Kelly for technical support throughout the experiment and analysis. 
This work was supported by JSPS KAKENHI Grant Numbers JP26247034, JP16H06288, and JP20H00149.
This work was supported in part by the RSF grant No.19-12-00325 and by the MHES (Russia) grant "Neutrino and astroparticle physics" No. 075-15-2020-778.
We acknowledge funding from the Spanish Ministerio de Economía y Competitividad (SEIDI-MINECO) under Grants No. PID2019-107564GB-I00. IFAE is partially funded by the CERCA program of the Generalitat de Catalunya.
We ackowledge the Swiss National Foundation through grant No. 200021 85012.
We further acknowledge the U.S.-Japan Science and Technology Cooperation Program in High Energy Physics, the support of the US Department of Energy, Office of High Energy Physics and the support from
the University of Pennsylvania.

\end{acknowledgments}

\bibliography{references}

\end{document}